\documentclass{pasa}%
\usepackage{graphicx}
\usepackage{latexsym}
\usepackage{amsmath}

\title[Stability of filaments in star-forming clouds and the formation of prestellar cores in them]{Stability of filaments in star-forming clouds and the formation of prestellar cores in them}
\author[Anathpindika. S., Freundlich, J.]{Anathpindika. S.$^{1}$\thanks{E-mail:
sumedh@physics.iisc.ernet.in} and  Freundlich, J$^{2}$\\
$^{1}$Indian Institute of Science, Bangalore 560012, India;
$^{2}$LERMA, Observatoire de Paris, CNRS, 61 av. de l'Observatoire, 75014 Paris, France\\ }
\jid{PASA}
\doi{10.1017/pas.\the\year.xxx}
\jyear{\the\year}

\begin{document}
\begin{abstract}
It is now widely accepted that dense filaments of molecular gas are integral to the process of stellar birth and potential star-forming cores often appear embedded within these filaments.
Although numerical simulations have largely succeeded in reproducing filamentary structure in dynamic environments such as in turbulent gas and while analytic calculations predict the formation of dense gas filaments via radial collapse, the exact process(es) that generate/s such filaments which then form prestellar cores within them, is unclear. In this work we therefore study numerically the formation of a dense filament using a relatively simple set-up of a uniform-density cylinder in pressure equilibrium with its confining medium. In particular, we examine if
its propensity to form a dense filament and further, to the  formation of prestellar cores within this filament bears on the gravitational state of the initial volume of gas. We report a radial collapse leading to the formation of a dense filamentary cloud is likely when the initial volume of gas is at least critically stable (characterised by the approximate equality between the mass line-density for this volume and its maximum value). Though self-gravitating, this volume of gas, however, is not seen to be in free-fall. This post-collapse filament then fragments along its length due to the growth of a Jeans-like instability to form prestellar cores like \emph{beads on a string}. We suggest, dense filaments in typical star-forming clouds classified as gravitationally super-critical under the assumption of : (i) isothermality when in fact, they are not, and (ii) extended radial profiles as against one that is pressure-truncated, thereby causing significant over-estimation of their mass line-density, are unlikely to experience gravitational free-fall. The radial density and temperature profile derived for this post-collapse filament is consistent with that deduced for typical filamentary clouds mapped in recent surveys of nearby star-forming regions. This profile is also in agreement with a Plummer-like density profile. For an isothermal filament though, the density profile is much steeper, consistent with the classic density profile suggested by Ostriker (1964). On the other hand, increasing the magnitude of the confining pressure such that the initial volume of gas is rendered gravitationally sub-critical is unable to collapse radially and tends to expand laterally which could possibly explain similar gas filaments found in recent surveys of some molecular clouds. Simulations were performed using smoothed particle hydrodynamics ({\small SPH}) and convergence of results is demonstrated by repeating them at higher resolution. Unlike some of the earlier work reported in literature, here we calculate gas temperature by solving the {\small SPH} energy equation and allow it to cool according to a cooling function.
\end{abstract}
\maketitle%
\section{INTRODUCTION }
\label{sec:intro}
Gas in Giant molecular clouds ({\small GMC}s) is distributed non-uniformly and appears to aggregate itself into isolated dense clumps or more contiguous and elongated, filament-like structures. Detailed observations of potential star-forming clouds  have demonstrated the ubiquitous nature of filamentary clouds. For example, giant molecular filaments on scales of a few parsecs have been reported in the inter-arm regions of the Milky-Way (Ragan \emph{et al.} 2014; Higuchi \emph{et al.} 2014 and other references therein).  On the other hand, relatively small (by about an order of magnitude compared to the former), dense filaments have also been reported within star-forming clouds in the local neighbourhood (e.g. Schneider \& Elmegreen 1979; Nutter \emph{et al.} 2008; Myers 2009; Andr{\' e} \emph{et. al.} 2010; Jackson \emph{et. al.} 2010; Arzoumanian \emph{et. al.} 2011; Kainulainen \emph{et al.} 2011, 2013 and Kirk \emph{et. al.} 2013 are only a few authors among an exhaustive number of them). 

Star-forming sites within {\small GMC}s are often found located within dense filamentary clouds or at the junctions of such clouds and further, these filamentary clouds usually show multiplicity, in other words, striations roughly orthogonal to the main filament, and form hubs (e.g. Palmeirim \emph{et al.} 2013; Hacar \emph{et al.} 2013; Schneider \emph{et al.} 2012 \& 2010; Myers 2009). In fact, inferences drawn from detailed observations of star-forming clouds have led some authors to suggest that turbulence-driven filaments could possibly represent the first phase in the episode of stellar-birth, followed by gravitational fragmentation of the densest filaments to form prestellar cores. Filamentary clouds therefore form a crucial part of the star-formation cycle. Consequently, a significant observational, theoretical and/or numerical effort has been directed towards understanding these somewhat peculiar clouds. In the last few years we have significantly improved our understanding about these clouds as a number of them have been studied in different wavebands of the infrared regime of the electromagnetic spectrum using sub-millimeter arrays on the {\small JCMT} and the Herschel (e.g. Nutter \& Ward-Thompson 2007; Andr{\' e} \emph{et al.} 2010; Men'schikov \emph{et al.} 2010; also see review by Andr{\' e} \emph{et al.} 2014). 

The stability and possible evolution of filamentary clouds has also been studied analytically  in the past and in more recent times. However, these models were usually developed under simplifying assumptions. For example, Ostriker (1964), developed one of the earliest models by approximating  a filamentary cloud as an infinite self-gravitating cylinder described by a polytropic equation of state and derived its density distribution.  In a later contribution, Bastien (1983), Bastien \emph{et al.} (1991) and Inutsuka \& Miyama (1992) studied the stability criteria of filamentary clouds under the assumption of isothermality and suggested that such clouds were more likely to form via the radial collapse of an initial cylindrical distribution of molecular gas. These models also demonstrated formation of prestellar cores along the dense axial filament  via Jeans-fragmentation. 

Formation of dense filaments via interaction between turbulent fluid flows has been demonstrated numerically by a number of authors (e.g. Klessen \emph{et al.} 2000; Bate \emph{et al.} 2003; Price \& Bate 2008, 2009; Federrath \emph{et al.} 2010a; Padoan \& Nordlund 2011 and Federrath \& Klessen 2012). Similarly in other recent contributions (e.g. Heitsch \emph{et al.} (2009); Peters \emph{et al.} 2012 and Heitsch 2013), respective authors specifically investigated the process that is likely to assemble a dense filament. The conclusion of these latter authors supports the idea of filament formation via radial collapse of gas followed by an accretional phase during which the filament acquires mass even as it continues to self-gravitate. In fact, Peters \emph{et al.} (2012) demonstrated the formation of filamentary clouds on the cosmic scale and argued that a filament was more likely to collapse radially and form stars along its length when confined by pressure of relatively small magnitude. In another recent contribution, Smith \emph{et al.} (2014), have demonstrated the formation of dense filaments in turbulent gas, however, they have not addressed the other crucial issue about the temperature profile of  these filaments; observations have revealed that gas in the interiors of dense filaments is cold at a temperature on the order of ~10 K (e.g. Arzoumanian \emph{et al.} 2010). Heitsch \emph{et al.} (2013) also address the issue of tidal-forces acting on the sides of a filamentary cloud having finite length and suggest that gas accreted at the ends can possibly give rise to fan-like features often found at the ends of infrared dark clouds ({\small IRDC}s). 

In a semi-analytic calculation, Jog(2013) demonstrated that tidal-forces also tend to raise the canonical Jeans mass, though compressional force-fields are more likely to raise the local density and therefore lower the Jeans mass. Freundlich \emph{et al.} (2014) derived a dispersion relation for a rotating non-magnetised filamentary cloud idealised as a polytropic cylinder with localised density perturbations. Under these simplifying assumptions, the authors demonstrated that the filament indeed developed Jeans-type instability with propensity to fragment on the scale of the local Jeans length.  These conclusions are in fact, consistent with those drawn in an earlier work by Pon \emph{et al.} (2011), or even those by Bastien \emph{et al.} (1991) who arrived at similar conclusions from their analytical treatment of the problem. 

On the other hand, Inutsuka \& Miyama (1997), showed that a cylindrical distribution of gas is unlikely to become self-gravitating as long as its mass per unit length was less than a certain critical value. Similar conclusions were also drawn by Fischera \& Martin (2012) when they performed a stability analysis of clouds idealised as isothermal cylinders. In the present work we study the dynamical stability of an initially non-self gravitating cylindrical distribution of molecular gas maintained initially in pressure equilibrium. In particular, we are interested to see how this distribution of gas behaves as it is allowed to cool in the presence of self-gravity. We would like to examine if a radial collapse does indeed ensue and the nature of the new state of equilibrium attained by the gas. Thus, we would like to investigate the physical conditions under which radial collapse is likely to be the favoured mode of evolution. Although the objective we have set ourselves is similar to that of Peters \emph{et al.} (2012), our work differs on three counts : (i) unlike Peters \emph{et al.} who assumed a polytropic equation of state to calculate gas temperature, we calculate gas temperature by solving the energy equation coupled with a cooling function, (ii) we do not use a filament that already is dynamically evolving instead, we begin by setting up a cylindrical cloud in pressure equilibrium and then examine if it does indeed collapse to form a thin dense filamentary cloud. This relatively simple set-up represents the intermediate stage of filament-formation, in other words, the stage after supersonically moving turbulent flows in a molecular cloud ({\small MC}) collide to form a filamentary distribution of warm gas (see e.g. Klessen \emph{et al.} 2000; Bate \emph{et al.} 2003; Price \& Bate 2008,2009; Federrath \emph{et al.} 2010a; Federrath \& Klessen 2012; Smith \emph{et al.} 2014), and more importantly, (iii) we investigate if gravitational supercriticality is a necessary pre-condition for filament-formation and follow the evolution of the dense filament further to study the formation of prestellar cores in it. 

 We observe that when the magnitude of the confining pressure is relatively small such that the initial volume of gas is at least critically stable, it does indeed collapse radially to assemble a thin dense filament which then fragments via a Jeans-like instability on the scale of the fastest growing unstable mode to form prestellar cores.  We also investigate physical properties such as the distribution of gas density and temperature within the post-collapse filament. We demonstrate that the density profile of the post-collapse isothermal filamentary cloud is indeed similar to the classic profile suggested by Ostriker (1964); the profile is Plummer-like for a filament that is allowed to cool dynamically. The article is divided as follows : in section 2 we discuss the stability of a pressure-confined isothermal cylinder. Then in section 3 we present the initial conditions used for this work and briefly describe the numerical algorithm used for the simulations. The results are presented and discussed in respectively, sections 4 and 5 before concluding in section 6.

\section{Stability of a pressure confined non-magnetic cylinder}
\label{sec: Stability of a pressure confined non-magnetic cylinder}
We consider a cylindrical distribution of isothermal molecular gas composed of the usual cosmic mixture (approximately 90\%  hydrogen and 10\% helium). The cylinder has initial radius, $R$, length, $L$, and maintained at temperature, $T_{gas}$. In a purely non-magnetic case gas within this cylinder is supported by thermal pressure, $P_{therm}$, against self-gravity and is confined externally by finite pressure exerted by the inter-cloud medium ({\small ICM}), $P_{icm}$ \footnote{By inter-cloud medium we mean here the medium that prevails between isolated clumps and filamentary clouds within a {\small GMC}.}. We assume an initial state of equilibrium so that the external pressure, $P_{icm}$, balances the internal pressure, $P_{therm}$. The internal and external pressure are both thermal by nature since particles representing the two media are not imparted any initial momentum.

The stability of a spherical gas body against self-gravity is defined in terms of the thermal Jeans mass. Now, for a cylindrical distribution of gas its mass line-density, $M_{l}(r)$, defined as
\begin{equation}
M_{l}(r=R) \equiv\ \frac{M(r=R)}{L} = \int_{0}^{R} dr\ 2\pi\ r\rho(r)\ =\ \pi R^{2}\rho_{gas}.
\end{equation}
This quantity is analogous to the mass of a spherical cloud and as will be seen below, is useful in describing the stability of a cylindrical cloud against self-gravity. Here, $\rho_{gas}$, is the initial density of the cylindrical distribution of gas, assumed to be uniform. A gas cylinder is stable against gravitational collapse as long as its mass line-density is smaller than the maximum value
\begin{equation}
(M_{l})_{max}\ =\ \frac{2a_{0}^{2}}{G},
\end{equation}
 where $a_{0}$ is the isothermal sound-speed for the cylinder. We define the stability factor, $S(r)$, as the ratio of the mass line-density for a cylinder against its maximum value. Thus,
\begin{equation}
S(r)\ =\ \frac{M_{l}(r)}{(M_{l}(r))_{max}}\ \equiv\ \frac{GM_{l}(r)}{2a_{0}^{2}};
\end{equation}
 When $S(R)$ exceeds unity, $M_{l}(R) > (M_{l})_{max}$, a radial collapse ensues and the cylinder may be described as \emph{super-critical}; cylinders with $S(R)$ less than unity are described as \emph{sub-critical} (Fischera \& Martin 2012). We describe cylinders with $S(R)\sim$ 1 as \emph{critically stable}.
As in our earlier work investigating the stability of starless cores (Anathpindika \& Di' Francesco 2013), in the following sections we will calculate the stability factor, $S(r)$, at different radii within the model cylindrical cloud over the course of its evolution to describe its stability against self-gravity. We now calculate the radius of the initial cylindrical distribution of gas. Keeping in mind that the pressure exerted by the confining {\small ICM} balances the thermal pressure within the cylinder, we can write
\begin{displaymath}
P_{icm}\ =\ \frac{a_{0}^{2}M_{gas}}{\pi R^{2}L}
\end{displaymath}
so that its radius,
\begin{equation}
R\ =\ \Big(\frac{a_{0}^{2}M_{gas}}{\pi L P_{ext}}\Big)^{0.5}.
\end{equation}

 Finally, following a simple dimensional analysis we define the free-fall time, $t_{ff}\equiv (GML^{-3})^{-1/2}$ $\sim$ $(G\rho)^{-1/2}$, which is slightly faster than the free-fall time in Jeans analysis;  $\rho\equiv\rho(r)$, is the density of the post-collapse filament having radius, $r < R$, and $a_{0}$, the sound-speed of gas within it. A free-fall ensues if $t_{ff}\ <\ t_{sc}$, where $t_{sc}$ is the sound-crossing time, $t_{sc}\equiv\frac{l}{a_{0}}$. Using the above inequality we obtain the condition for Jeans instability as, 
\begin{equation}
l\geq L_{Jeans}\sim \frac{a_{0}}{(G\rho)^{1/2}}.
\end{equation}
The corresponding mass of the fragment is $M_{frag}\sim M_{l}L_{Jeans}$, which, using Eqn. (3) above, can be re-cast as 
\begin{equation}
M_{frag}\ =\ \frac{2a_{0}P(r=r_{flat})S(r=r_{flat})}{(G\rho)^{3/2}},
\end{equation}
$r_{flat}$, being the radius of the filament over which gas density remains approximately uniform before turning over in its wings.

\begin{table*}
\begin{minipage}[h]{0.9\textwidth}
\caption{Parameters used for each realisation.}             
\label{table:1}      
\centering          
\renewcommand{\footnoterule}{}  
\begin{tabular}{r l l r l l l l l l l l r}   
\hline\hline     
Serial & $\Big(\frac{P_{ext}}{k_{B}}\Big)$\footnote{Magnitude of the external confining pressure that balances the pressure within the cylinder, $P_{therm}$. } & M$_{gas}$ & $R$\footnote{Calculated using Eqn. (4) in the text.} & $(\mathrm{M}_{l})_{init}$\footnote{Initial mass-per unit length of the filament, i.e., the mass line-density.} & T$_{gas}$\footnote{The initial gas temperature.} & $(\mathrm{M}_{l})_{max}$\footnote{Maximum mass line-density defined by Eqn. (2).}  & Gas& N$_{tot}$\footnote{Total number of particles including those representing the {\small ICM }.} & N$_{gas}$\footnote{Number of gas particles} & $\chi$\footnote{Resolution factor} & Perturbations\footnote{If whether perturbations described in \S 3.1 were imposed in a particular realisation.}\\
No. & $[\times 10^{4}\mathrm{K} \mathrm{cm}^{-3}]$ & [M$_{\odot}$] & [pc] & $[M_{\odot}\mathrm{pc}^{-1}]$ & [K] & & Dynamics & \\
\hline
1 & 1.5 & 210 & 0.589 & 42 & 25 & 41.37 & ISOTHERMAL & 478884 & 339941 & 1.6 & N \\
2 & 1.5 & 244.7 & 0.636 & 48.94 & 25 & 41.37 & Cooling Func. & 478884 & 339941 & 1.6 & N\\
3 & 2.4 & 244.7 & 0.636 & 48.94 & 40 & 66.19 & Cooling Func. & 478884 & 339941 & 1.6 & N\\
4 & Same as 2 & & & & & & & & & & Y \\ 
& & & & & & & & & & & {\small (Fundamental} \\
& & & & & & & & & & & {\small mode)} \\
5 &  Same as 2 & & & & & & & & & & Y \\ 
& & & & & & & & & & & {\small (Second } \\
& & & & & & & & & & & {\small Harmonic)} \\
6 & Same as 4 & & & & & & &1545929 & 765378 & 2 & Y \\
7 & Same as 5 & & & & & & &1545929 & 765378 & 2 & Y \\
8 & Same as 3 & & & & & & &1545929 & 765378 & 2 & N \\
\hline                    
\hline                  
\end{tabular}
\end{minipage}
\end{table*}

\section{Physical details}
\subsection{Initial conditions}
In this demonstrative work we began by placing a non-self gravitating uniform-density cylindrical distribution of molecular gas in pressure equilibrium with its surrounding medium. To this initial distribution of gas we assigned a feducial temperature (see Table 1). In this work the gas pressure and pressure exerted by the external medium are purely of thermal nature. While ambient physical conditions such as the turbulent Mach number in a {\small MC} could possibly influence filament properties, we leave this aspect of the problem for future investigation. Other free parameters in the set-up were, the length of the cylinder, $L$= 5 pc, contrast, $D$=10, between the initial gas density within the cylinder and its surrounding medium, the average initial gas density within the cylinder, $\bar{n}_{gas}$ = 600 cm$^{-3}$ and the initial mass line-density, $(M_{l})_{init}$. This choice of gas density is also consistent with densities found in a typical star-forming cloud (e.g. Andr{\' e} \emph{et al.} 2013). The temperature of the confining medium is calculated by using the condition of pressure-balance at the surface of the cylinder. The initial radius of the cylinder was calculated using Eqn. (4) above. Though simple, this set of initial conditions is suitable for the extant purposes of studying the origin of the density profile and the temperature profile of a typical filament and the formation of prestellar cores along the length of this filament. Parameters used in the realisations performed in this work have been listed in Table 1. Sinusoidal density perturbations along the length of the initial distribution of gas were imposed in five simulations; see Table 1. These perturbations were imposed by revising particle positions, $x_{p}$, to $x_{p}'$, such that the following expression was satisfied,
\begin{displaymath}
x_{p} = x_{p}' + \frac{A L_{Jeans}}{2\pi}\sin(kx_{p}')
\end{displaymath}
(Hubber \emph{et al.} 2006), where the wave-number, $k = \frac{n\pi}{L_{Jeans}}$, the integer, $n$ = 1, 2 for respectively, the fundamental mode and the second harmonic; $A$ = 0.1, is the amplitude of perturbations.

The number of gas particles in a simulation are calculated according to -
\begin{displaymath}
\mathrm{N}_{gas} = \frac{1}{8}\cdot\frac{3}{4\pi}\cdot \mathrm{N}_{neibs}\cdot\Big(\frac{V_{init}}{h_{avg(init)}^{3}}\Big),
\end{displaymath}
where, $V_{init}$, is the volume of the initial distribution of gas, while the initial average smoothing length, $h_{avg(init)}$, 
was set equal to $0.25L_{Jeans}/\chi$; $\chi$ being the resolution parameter listed in column 11 of Table 1 and defined by Eqn. (10) below; the Jeans length, $L_{Jeans}\sim 0.08$ pc, at the density threshold($\sim 10^{-19}$ g cm$^{-3}$), adopted for representing prestellar cores in this work at a temperature of 10 K. Particles representing the initial distribution of gas were assembled by extracting a cylinder of unit size and length from a pre-settled glass like distribution. This unit-cylinder was then stretched to the desired dimensions and then placed in a confining medium of external-pressure, modelled using the {\small ICM} particles. The envelope of {\small ICM} particles had a thickness, 30$h_{avg(init)}$, along each spatial dimension of the cylinder. The {\small ICM} particles are special {\small SPH} particles that exert only hydrodynamic force on the normal gas particles; gas particles on the other hand, exert both, gravitational and hydrodynamic forces. The cylinder and its confining medium were then placed in a box lined with boundary particles that prevent {\small SPH} particles from escaping. Unlike the gas and the {\small ICM} particles, the boundary particles are dead and do not contribute to {\small SPH} forces. The layer of dead boundary particles was there to simply hold the gas+{\small ICM} particles in the box and prevent them from diffusing away. Particles within the set-up were initially stationary as no external velocity field was introduced.

\subsection{Numerical method : Smoothed Particle Hydrodynamics (SPH)}
Simulations were performed using our well tested {\small SPH} algorithm, {\small SEREN}, in its conservative energy mode (Hubber \emph{et al.} 2011). The {\small SPH} calculations in this work also include contributions due to the artificial conductivity prescribed by Price (2008), to avoid any spurious numerical artefacts possible due to the creation of a gap between multi-phase fluids. An {\small SPH} particle, in this work, had a fixed number of neighbours, $N_{neibs}$ = 50. Gas cooling was implemented by employing the cooling function given by the equation below, 
\begin{equation}
\frac{\Lambda(T_{gas})}{\Gamma_{CR}} = 10^{7}\textrm{exp}\Big(\frac{-1.184\times 10^{5}}{T_{gas}+1000}\Big) + \nonumber
\end{equation}
\begin{equation}
 1.4\times 10^{-2}\sqrt{T_{gas}}\Big(\frac{-92}{T_{gas}}\Big)\ \textrm{cm}^{3}
\end{equation}
(Koyama \& Inutsuka 2002).
Constant background heating due to cosmic rays is provided by the heating function, $\Gamma_{CR} = 2\times 10^{-26}$ erg s$^{-1}$. The temperature of an {\small SPH} particle is calculated by revising its internal energy. The equilibrium temperature, $T_{eq}$,  attained by \textbf{a} small {\small SPH} particle corresponds to equilibrium energy, $u_{eq}$, and the timescale, $\tau$, over which energy is either radiated or acquired by a particle is,
\begin{equation}
\tau\ =\ \Big\vert \frac{u_{p} - (u_{p})_{eq}}{(n^{2}\Lambda - n\Gamma_{CR})}\Big\vert.
\end{equation}
Then, after a time-step $dt$, the energy $u$ of a particle is revised  according to,
\begin{equation}
u = u_{eq} + (u - u_{eq})\textrm{exp}(-dt/\tau)
\end{equation} 
(V{\' a}zquez-Semadeni \emph{et al.} 2007). However, a more accurate calculation of the gas temperature demands solving the radiative transfer equation(e.g. Krumholz \emph{et al.} 2007; Price \& Bate 2009; Bate 2012), and accounting for the molecular chemistry in the cloud (Glover \emph{et al.} 2010). We leave this for a future work.\\
\emph{The {\small SPH} sink particle} \\
 We represent a prestellar core by an {\small SPH} sink particle in this work, for it is not our intention to follow the internal dynamics of a core. An {\small SPH} particle is replaced with a sink only when the particle satisfies the default criteria  for sink-formation which include, negative divergence of velocity and acceleration of the seed particle (see Bate, Bonnell \& Price 1995; Bate \& Burkert 1997 and Federrath \emph{et al.} (2010b)). We set the density threshold for sink-formation at  $\rho_{thresh}\sim 10^{-19}$ g cm$^{-3}$ ($\sim 10^{5}$ cm$^{-3}$), which despite being on the lower side, is consistent with the average density of a typical prestellar core. The interested reader is referred to Hubber \emph{et al.} (2011) for other technical details related to implementation of this prescription in our algorithm SEREN. The radius of a sink particle was set at, $r_{sink}\sim\ 2.5h_{p}$, where $h_{p}$ is the smoothing length of the {\small SPH} particle that seeds a sink. Sink particles in the simulation therefore have different radii. A sink interacts with normal gas particles only via gravitational interaction and accretes gas particles that come within the pre-defined sink-accretion radius, $r_{sink}$.\\ \\

\begin{figure*}
\centering
      \includegraphics[angle=270,width=\textwidth]{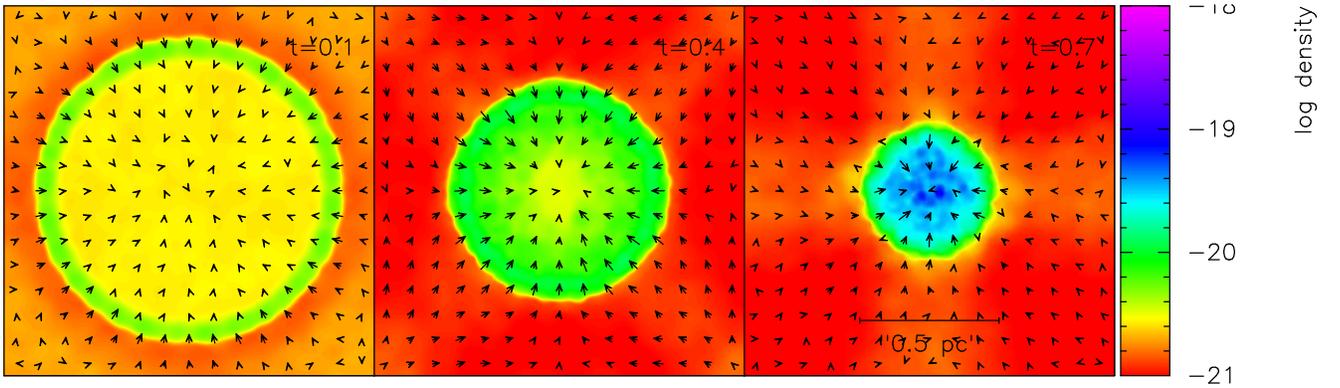}
      \caption{Cross-sectional rendered density-plots of the mid-plane of the collapsing cylinder in realisation 1. Overlaid on these images are velocity vectors representing local gas-motion; shorter vectors indicate relatively lower velocity as compared with those that are longer. Time in units of Myr has been marked on the top-right hand corner of each image.}
\end{figure*}

\subsection{Resolution}
The average initial smoothing length, $h_{avg(init)}$, for the ensemble of calculations discussed in this article is calculated as -
\begin{displaymath}
h_{avg(init)}\ =\ \Big(\frac{3}{32\pi}\cdot\frac{N_{neibs}}{N_{gas}}\cdot V_{init}\Big)^{1/3},
\end{displaymath}
where the number of neighbours, $N_{Neibs}$ = 50.
The diameter of an {\small SPH} particle, $d\ =\ 4h_{avg(init)}$, so that the resolution criterion becomes,
\begin{equation}
\mathcal{X}\ =\ \frac{\L_{Jeans}}{d};
\end{equation}
$\mathcal{X}\ \geq\ 1$, and $L_{Jeans}$, the typical length-scale of fragmentation has been defined by Eqn. (5) above.  Evidently, higher the value of $\mathcal{X}$, better the resolution. Value of the resolution factor, $\mathcal{X}$, for each simulation has been listed in column 10 of Table 1. The resolution criterion defined by Eqn. (10) above is the {\small SPH} equivalent of the Truelove criterion (Truelove \emph{et al.} 1997), defined originally for the grid-based Adaptive Mesh Refinement algorithm. It has been shown that the gravitational instability is reasonably well-resolved for $\chi$ = 1, and need be $\sim$2, for a good resolution of this instability. Note that some authors prefer to define the resolution factor, $\mathcal{X}$, as the inverse of the fraction on the right-hand side of Eqn. (10), in which case the criterion for good resolution  becomes, $\mathcal{X}\ \leq\ 1$ (e.g. Hubber \emph{et al.} 2006). All our simulations discussed in this work satisfy this resolution criterion. 

However, this choice of resolution is reportedly insufficient to achieve convergence in calculations of crucial physical parameters such as the energy and momentum, but sufficient to prevent artificial fragmentation of collapsing gas (Federrath \emph{et al.} 2014). In order to alleviate this latter problem these authors suggest a more stringent resolution criterion, $\chi\gtrsim$ 15. However, our extant purpose here is limited to studying the stability of a filamentary cloud and its fragmentation so that the adopted choice of numerical resolution($\chi\sim$ 2), should be sufficient. In fact, in \S 4.1 below we demonstrate numerical convergence over the observed fragmentation of the post-collapse filament. Finally, the sink radius, $r_{sink}$, defined in \S 3.2 above is $\sim 0.08$ pc in simulations developed with the least number of particles and $\sim 0.05$ pc in Cases 6 and 7, those developed with the highest resolution in this work. Although prestellar cores of smaller size are known to exist in star-forming clouds (see for e.g. Jijina \emph{et al.} 1999), the resolution adopted here is sufficient for a demonstrative calculation of the kind presented in this work. The typical width of the post-collapse filament, $\sim$0.1 pc, is also consistently resolved by a minimum of 80 particles along its  length.

Since this volume of gas is allowed to cool, what follows, is the fragmentation of cooling gas. The length-scale of this fragmentation is comparable to the thermal Jeans length (see for e.g. V{\' a}zquez-Semadeni \emph{et al.} 2007), which is much larger than the initial width of the cylinder so that there is no fragmentation in the radial direction, as will be demonstrated below. Furthermore, we have set our resolution to the Jeans length, $L_{Jeans}$, evaluated at 10 K, which is lower than the average gas temperature within interiors of the filament as will be evident in the following section. Our simulations therefore have the desired spatial resolution to represent core-formation in the post-collapse filament. 

 \begin{figure}
   \centering
   \mbox[\includegraphics[angle=270,width=0.5\textwidth]{case1_rend_sinks.eps}
        \includegraphics[angle=270,width=0.5\textwidth]{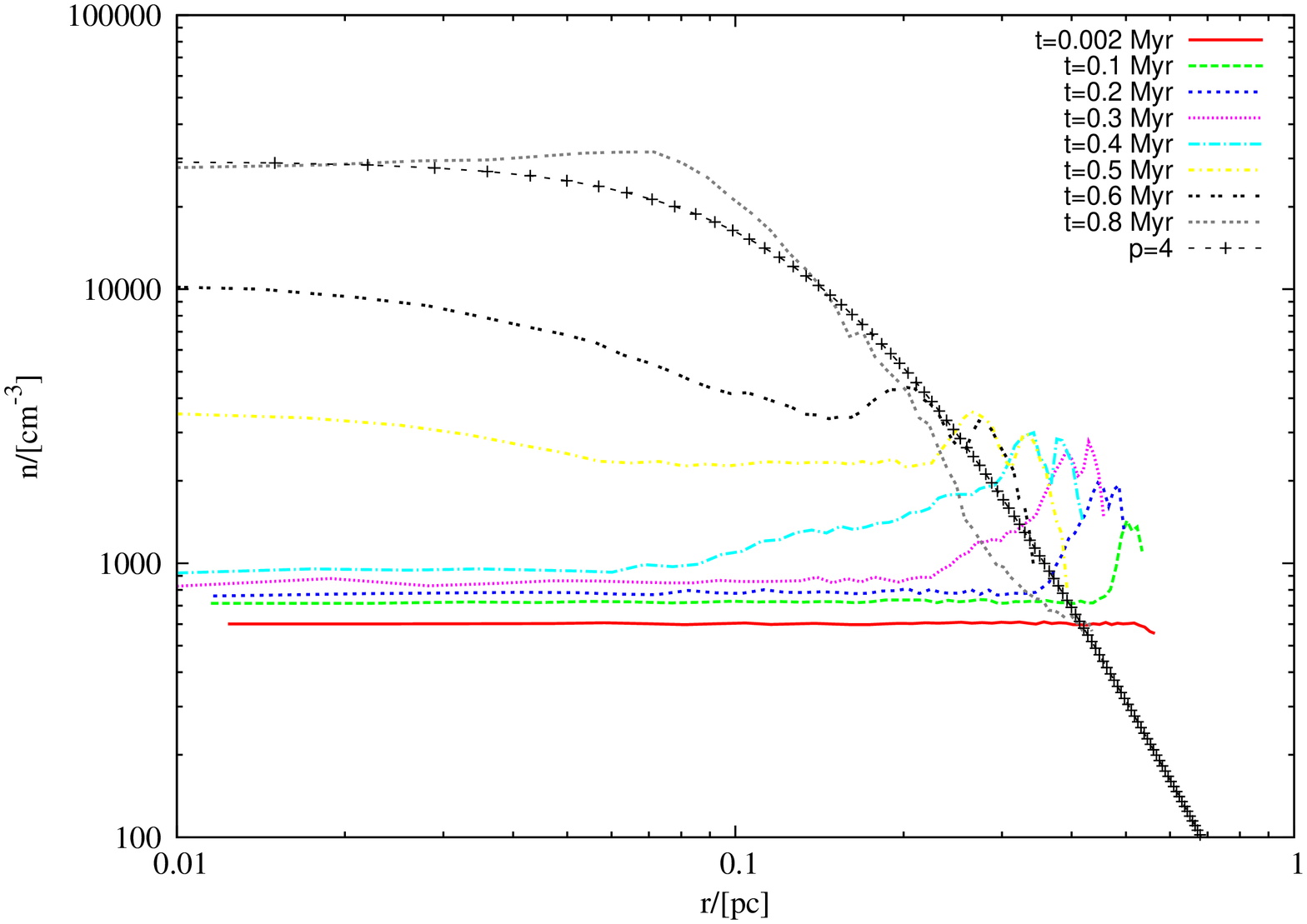}]
      \caption{\emph{Upper-panel} A rendered density plot through the mid-plane of the post-collapse filament in realisation 1. Black blobs on this plot represent the positions of prestellar cores in this filament and those that have formed at this epoch are separated by approximately a Jeans-length for this calculation ($\sim$0.2 pc). \emph{Lower-panel} Plot showing a time-sequence of density profile within the collapsing filament for this case. The density distribution for the post-collapse filament ($t=0.8$ Myr), is very well approximated by the Ostriker profile.}
         \label{FigVibStab}
   \end{figure}

 \begin{figure}
   \centering
   \includegraphics[angle=270,width=0.5\textwidth]{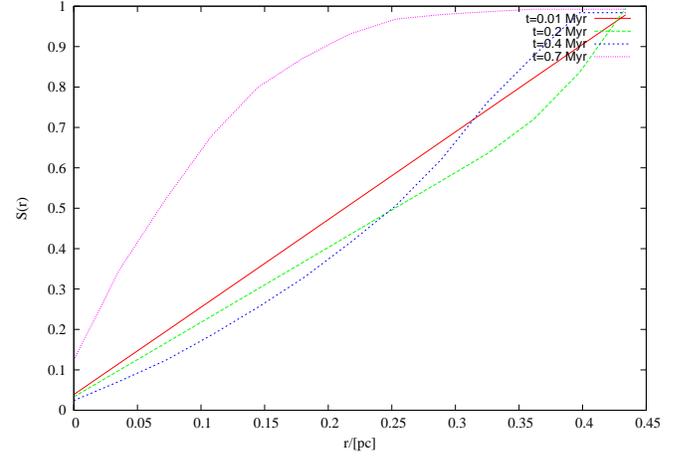}
      \caption{Shown here is the time-variation of the stability factor, S(r), defined by Eqn. (3) as a function of the radial coordinate within the collapsing filament in realisation 1. Interestingly, gas in the collapsing cylinder always remains gravitationally sub-critical; $S(r) < 1$.}
         \label{FigVibStab}
   \end{figure}

\begin{figure}
   \centering
   \includegraphics[angle=270,width=0.5\textwidth]{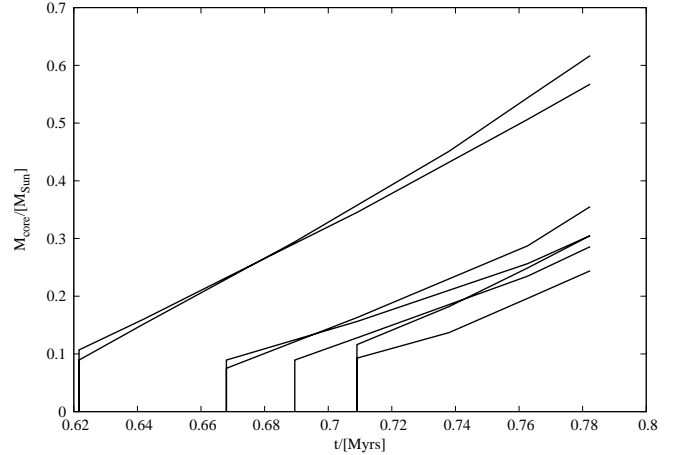}
      \caption{A plot showing the accretion history of the cores that form in the post-collapse filament in the first realisation.}
         \label{FigVibStab}
   \end{figure}

\section{Results}
\label{sec: Results}
Our interest lies particularly in examining the possibility of forming a dense filamentary cloud out of a cylindrical volume of gas confined by thermal pressure. This possibility is explored for 
different choices of initial conditions; first, when the volume of gas is initially critically stable, then marginally super-critical and finally, sub-critical. Interestingly though, the initial cylindrical distribution of gas evolves radically differently
 if it is at/above critical stability as compared to when it is sub-critical. We also  vary the spatial resolution and compare results from simulations developed by using a cooling function with that from an isothermal calculation.

\subsection{Case 1 (Serial nos. 1, 2, 4, 5, 6 and 7 in Table 1; $\frac{P_{icm}}{k_{B}}$ = 1.5$\times\ 10^{4}$ K cm$^{-3}$)}
The initial cylindrical distribution of gas, in all these test cases, has comparable mass and an identical magnitude of confining pressure, $P_{icm}$. One of these, test realisation 1, was developed under the assumption of isothermality while the cooling function defined by Eqn. (7) above, was used in the remaining test cases. The mass line-density, $(M_{l})_{max}$, defined by Eqn. (2) is slightly less than that for the initial distribution of gas, $(M_{l})_{init}$, for all the test realisations, except for the first where the two are of comparable magnitudes. The initial distribution of  gas is therefore marginally super-critical in all realisations; gas in the first realisation is approximately in  equilibrium in the sense, $S(R)\sim 1$ (see Table 1), $R$ being the radius of the cylinder. In all these realisations the radius of the initial cylindrical distribution of gas begins to shrink due to the onset of an inwardly propagating compressional wave. This radial contraction of the initial cylindrical distribution in outside-in fashion is visible in the rendered density images of the cross-section through the mid-plane of the collapsing cylinder shown in Fig. 1. Overlaid on these plots are the velocity vectors that denote the direction of local gas-motion.

This contraction soon assembles a thin dense filament along the axis of the cylinder and prestellar cores, represented by {\small SPH} sink particles, begin to form along the length of the filament. At this point a word of caution would be appropriate. By a compressional wave we do not imply that the external pressure compresses the initial cylindrical distribution of gas in the radial direction. That hydrostatic cylinders cannot be compressed in this manner has been demonstrated by Fiege \& Pudritz (2000). In this work density enhancement along the cylinder-axis is the result of gas being squeezed during the radial collapse of the initial configuration. We reserve the attribute of a compressional wave to describe this process. Shown on the upper-panel of Fig. 2 is a rendered density plot of the post-collapse filament in realisation 1. The black blobs on this image represent the location of prestellar cores that begin to form along the length of this filament. 

Evidently, the cores that had formed in this filament at the time of terminating the calculation are separated by approximately a Jeans length($\sim$ 0.2 pc at $T_{gas}$ = 25 K; avg. density $\sim 10^{-19}$ g cm$^{-3}$), defined by Eqn. (5). The image on the lower-panel of Fig. 2 is a plot of the radial distribution of gas density within the collapsing cylindrical cloud in this realisation. This plot and all other similar plots in this paper were generated by taking a transverse section through the mid-plane of the filament.
The inwardly propagating compressional disturbance is evident from the density plots at early epochs of the collapse. As the gas is steadily accumulated in the central plane of the cylinder, the density of gas there rises and eventually develops a profile that matches very well with the one suggested by Ostriker (1964),
\begin{equation}
 \rho(r) = \frac{\rho_{c}}{\Big[1 + \Big(\frac{r}{r_{flat}}\Big)^{2}\Big]^{p/2}};
\end{equation}
with $p$ = 4. Here $\rho_{c}\equiv\rho(r=r_{flat})$; $r_{flat}$ being the radius at which the density profile develops a knee. The density of this filament falls-off relatively steeply in the outer regions which is consistent  with that reported by  Malinen \emph{et al.} (2012)  for typical filamentary clouds in the Taurus {\small MC}. Furthermore, this figure also shows, $r_{flat}\sim$ 0.1 pc, which is comparable to the radius of typical filaments found in eight nearby star-forming clouds in the Gould-Belt (e.g. Arzoumanian \emph{et al}. 2011; Andre \emph{et al.} 2014). 

Another interesting conclusion that can be drawn from this realisation is that about the stability of a filamentary cloud. It can be seen from Fig. 2 that the centrally assembled, post-collapse dense filament at the epoch when calculations for this realisation were terminated, has a radius on the order of 0.1 pc, the point at which the density-profile develops a knee. An average density of $\sim 10^{-19}$ g cm$^{-3}$, the threshold for core-formation in these calculations, would suggest a mass line-density, $M_{l}\sim$ 50 M$_{\odot}$ pc$^{-1}$, for this post-collapse filament which is significantly higher than, $(M_{l}(T_{gas}=25 K))_{max}\sim$ 41 M$_{\odot}$ pc$^{-1}$. The filament would therefore be expected to be in free-fall and should collapse rapidly to form a singular line. However, such is not to be the case  and Fig. 3 shows that the collapsing cylindrical cloud, or indeed the post-collapse filament, in fact, remains sub-critical at all radii, $r<R$. We will discuss later in \S 5.1 the implications of this finding for the classification of observed filamentary clouds in typical star-forming regions.

In Fig. 3 we have shown the radial variation of the stability factor, $S(r)$, defined by Eqn. (3), and integrated over the radius of the collapsing cylindrical cloud. For the purpose, using the Sturges criterion for optimal bin-size (Sturges 1926), we divided the cylindrical distribution of gas into a number of concentric cylinders having incremental radii in the radially outward direction. Number of these concentric cylinders is given by, $N_{cyl} = (1 + \ln(N_{gas}))$; $N_{gas}$ being the number of gas particles in a realisation. The innermost cylinder would therefore have the smallest radius and radii of successive cylinders in the outward direction would be incremented by a small increment, $dr$. Thus, if $r_{i}$ is the radius of the $i^{th}$ cylinder, then the radius of the $(i+1)^{th}$ cylinder is simply, $r_{i+1} = r_{i} + dr$; $dr\ll R$. The initial distribution of gas in this case was characterised by $S(R)\sim 1$. 

Evident from the plot shown in Fig. 3 is the fact that the gravitational state of the collapsing cylinder does not vary during the process. Within the cylinder gas always remains gravitationally sub-critical so that, $S(r<R) < 1$, although it does rise steadily close to the centre where the density is the highest in the post-collapse filament. In other words, shells of gas within the collapsing cylinder though self-gravitating, are not essentially in free-fall. That this must be the case is also evident from the density plots shown in Fig. 1. From these plots it can be seen that gas inside the collapsing cylinder has a relatively small velocity in comparison with that in the outer regions that is pushed in by the compressional wave. Also, the observed outside-in type of collapse lends greater credence to the conclusion that gas within the collapsing cylinder must be gravitationally sub-critical, for had it been super-critical, the collapse would have been of the inside-out type, i.e., the innermost regions of the cylinder would have collapsed rapidly followed by the outer regions.

We had presented a similar argument in an earlier work (Anathpindika \& di' Francesco 2013), where hydrodynamic models developed to explain why some prestellar cores failed to form stars despite having masses well in excess of their respective Jeans mass, were discussed. In that work it was demonstrated that cores could indeed contract in the radial direction without becoming singular (i.e., a free-fall collapse was not seen in these cores), if the mass of infalling gas at radii within the core was smaller than the local Jeans mass, or equivalently, we had suggested that the Jeans condition had to be satisfied at all radii within a collapsing core for it to become singular (protostellar). In the present work we have used the mass line-density instead of the thermal Jeans mass to discuss the stability of a cylindrical volume of gas. We observe, the collapsing cylinder appears to evolve through quasi-equilibrium configurations where thermal pressure provides support against self-gravity and a radial free-fall does not ensue. Interestingly, we observe that prestellar cores can indeed form in a filamentary cloud that is not in radial free-fall. These cores acquire mass by accreting gas and shown in Fig. 4 is the accretion history of these cores.

\begin{figure}
   \centering
   \includegraphics[angle=270,width=0.5\textwidth]{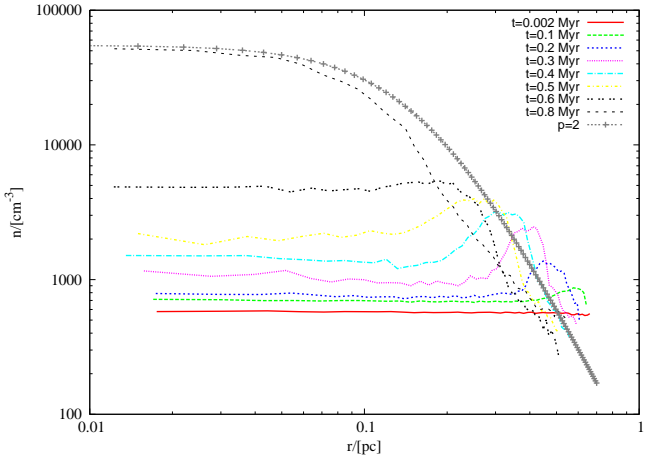}
      \caption{As in the plot shown in Fig. 1, this is the radial distribution of gas density within the collapsing filament at different epochs in realisation 2. }
         \label{FigVibStab}
   \end{figure}

\begin{figure}
   \centering
   \includegraphics[angle=270,width=0.5\textwidth]{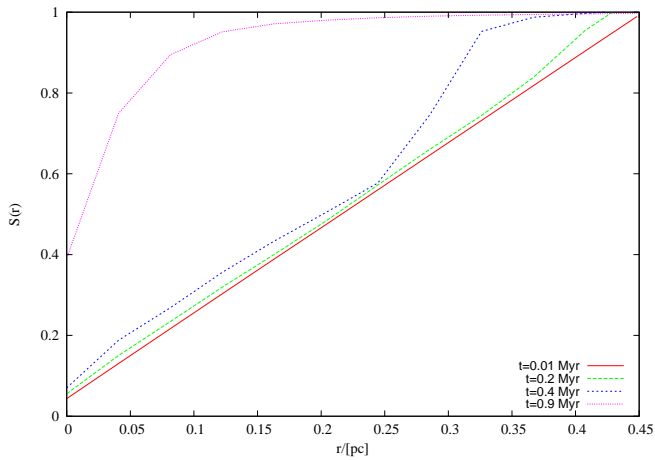}
      \caption{As in Fig. 2, shown in this plot is the time-variation of the stability factor, $S(r)$, calculated for different radii at various epochs of the collapsing filament in realisation 2.}
         \label{FigVibStab}
   \end{figure}

\begin{figure}
   \centering
   \includegraphics[angle=270,width=0.5\textwidth]{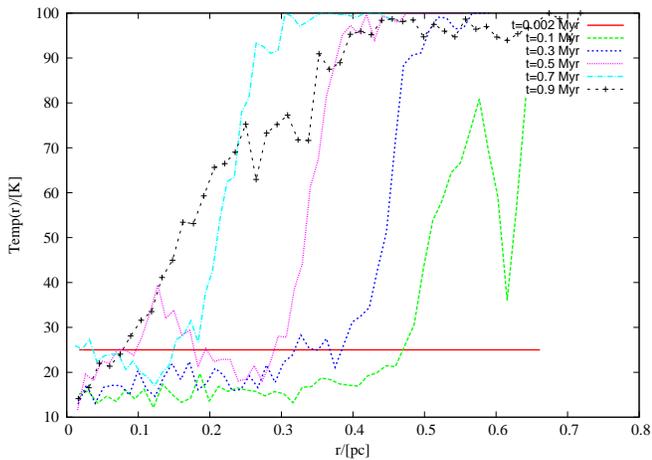}
      \caption{This plot shows the radial variation of density averaged gas temperature at different epochs within the collapsing filament in realisation 2.}
         \label{FigVibStab}
   \end{figure}

\begin{figure}
   \centering
   \includegraphics[angle=270,width=0.5\textwidth]{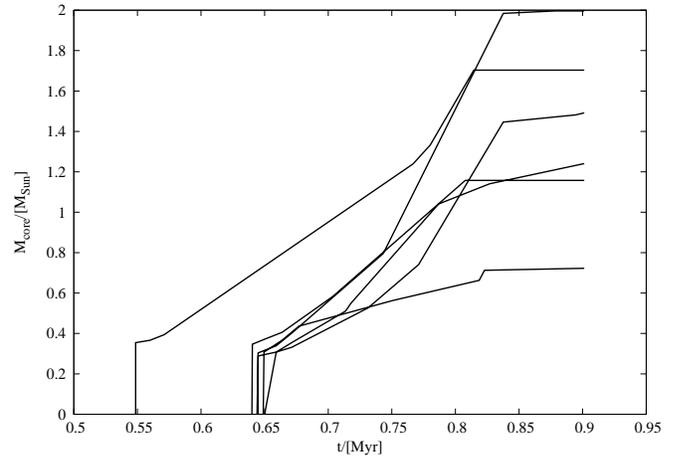}
      \caption{Same as the plot in Fig. 3, but now for the cores that form in the post-collapse filament in realisation 2.}
         \label{FigVibStab}
   \end{figure}

Realisation 2 is a repetition of the previous simulation but the gas temperature was now calculated by solving the energy equation and further, the energy of a gas particle was corrected according to Eqn. (9) to account for gas cooling/heating. As with the isothermal cylindrical  cloud in the previous realisation, the marginally super-critical initial distribution of gas in this case also begins to collapse in the radial direction. And, as in the previous case the inwardly propagating compressional wave assembles a dense filament of gas along the axis of the initial cylindrical cloud. Shown in Fig. 5 is the radial distribution of gas density within the collapsing cylindrical cloud which, though remarkably similar to that shown in Fig. 2 for the isothermal filament, is relatively shallow (Plummer-like; $p$ = 2 in Eqn. (11)). The radial variation of the stability factor, $S(r)$, within the collapsing cylinder for this realisation has been shown in Fig. 6. This plot is qualitatively similar to the one produced for simulation 1 and shown earlier in Fig. 3. As with this former plot, gas within the collapsing cylinder in this case also always remains less than unity, $S(r<R)<1$, suggesting that gas is not in free-fall. Also, the filament in this case is assembled on a timescale comparable to that in the isothermal realisation. This is probably because, despite the gas-cooling, the collapsing gas is still gravitationally sub-critical. A direct comparison of the mass line-density, $M_{l}(R)$, against the threshold for stability, $(M_{l})_{max}$, however, suggests otherwise. This is because, implicit in a comparison of this kind is the assumption of isothermality and uniform distribution of gas density within the post-collapse filament. 

On the contrary, and as is evident from Fig. 7, the gas temperature within the post-collapse filament is hardly uniform. The cold central region of the filament is cocooned by the relatively warm jacket of gas. The innermost regions of this filament acquire an average temperature of ~10 K which is consistent with that reported from observations of filamentary clouds in typical star-forming regions (see for e.g. Palmeirim \emph{et al.} 2013; Arzoumanian \emph{et al.} 2011; Andr{\' e} \emph{et al.} 2010). Finally, shown in Fig. 8 is the history of core-formation in this filament. It is qualitatively similar to that for the isothermal filament plotted in Fig. 4. To this end, following the formation of the first core, the next set of cores form relatively quickly. Also, the filament in the second realisation, where gas was allowed to cool, begins to form cores earlier than the isothermal filament and in general, cores in the latter are somewhat less massive than those in the former.  This is the result of a higher mass line-density of the cooling filament relative to the isothermal filament in the earlier realisation.

\begin{figure}
   \centering
   \includegraphics[angle=270,width=0.5\textwidth]{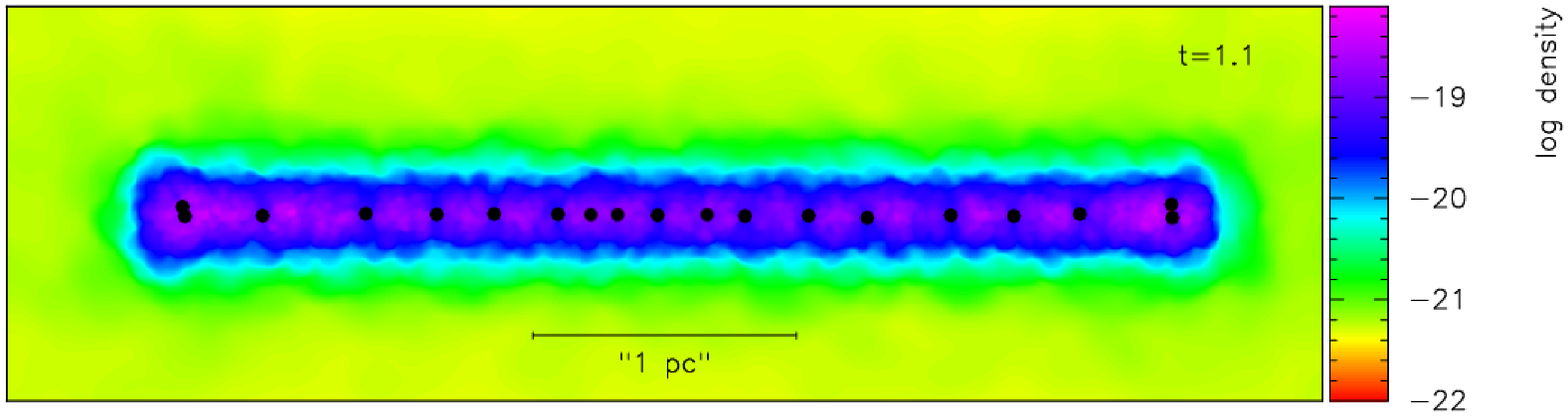}
   \includegraphics[angle=270,width=0.5\textwidth]{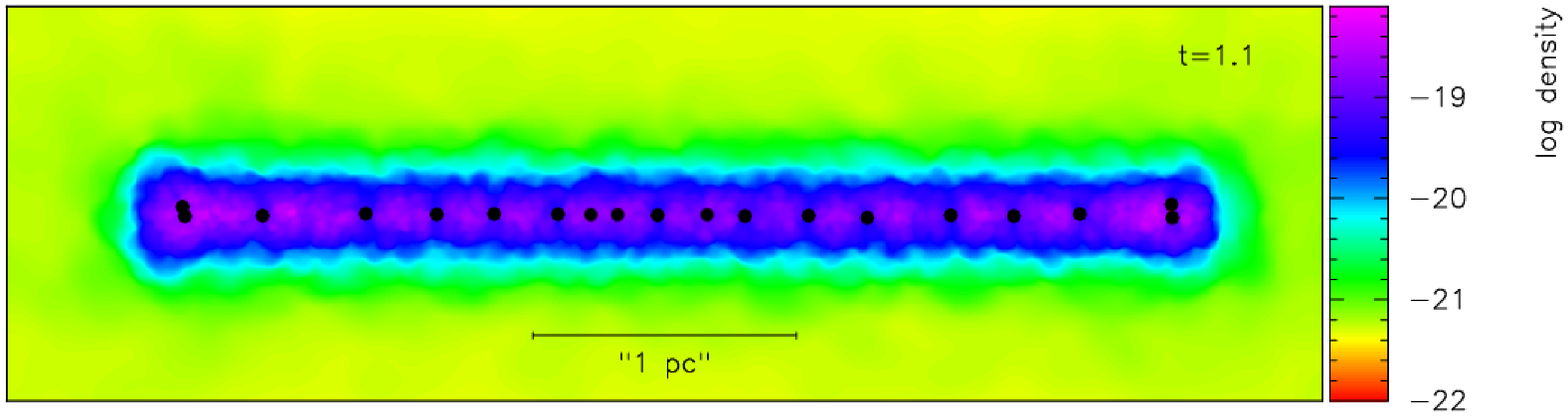} 
      \caption{\emph{Upper-panel:} Rendered image showing a projection of the mid-plane of the post-collapse filament($t$=1.) in realisation 6 (\emph{only a small volume of the medium confining the filament has been shown on this plot}). Fine black blobs on top of the image represent the positions of cores in the post-collapse filament at the time of termination of calculations. Perturbations, in this realisation, were imposed on the length-scale $L_{Jeans}$, defined by Eqn. (5). Not all density perturbations have condensed at this epoch, but those that are relatively closely spaced are separated on a scale on the order of $L_{Jeans}\sim 0.08$ pc.\emph{Lower-panel:} Same as the picture shown in the upper- panel, but now for the realisation 4 that was developed with a somewhat lower resolution, though sufficient to avoid artificial fragmentation along the length of the post-collapse filament.}
         \label{FigVibStab}
   \end{figure}

\begin{figure}
   \centering
   \includegraphics[angle=270,width=0.5\textwidth]{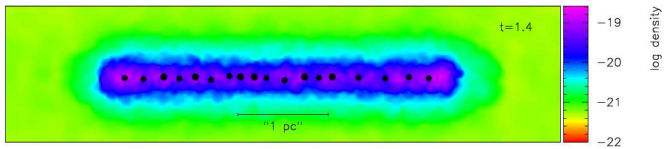}
      \caption{Similar to the plot shown in Fig. 8, but with perturbations now imposed on the length-scale, $L_{Jeans}/2$ (the second harmonic), in realisation 7. Again, not all density perturbations have condensed at this epoch, but those that are relatively closely spaced are separated on a scale on the order of $L_{Jeans}/2\sim 0.04$ pc. }
         \label{FigVibStab}
   \end{figure}
In realisation 1, for instance, where no external perturbations were imposed, we observed that the separation between 
cores in the post-collapse filament was on the order of the corresponding Jeans length. Next, we use the same set of initial conditions as those for realisation 2 in this next set of four realisations namely, 4, 5, 6 and 7, but now by imposing perturbations on the initial distribution of gas as described in \S 3.1. Perturbations for realisation 4 were imposed on the scale of the fragmentation length defined by Eqn. (5), which at 10 K and average density, $\sim 10^{-19}$ g cm$^{-3}$, the threshold for core-formation, is $L_{Jeans}\sim$ 0.08 pc.  We define this as the fundamental mode of perturbation. In the next realisation, numbered 5 in Table 1, we imposed the second harmonic of this perturbation. We repeat this set of two calculations with a higher number of particles, labelled 6 and 7 respectively, in Table 1. As with the gas in realisation 2 that was subject to cooling, in this case also, we observe that the initial distribution of gas collapsed radially to form a thin filament aligned with the central axis of the cylinder. Shown in Figs. 9 and 10 are the rendered images of the post-collapse filaments that form in realisations 6 and 7, those that have the best resolution in this set of calculations. For comparison purposes we have also shown a plot of the post-collapse filament from realisation 4 on the lower-panel of Fig. 9. The physical parameters for this realisation are the same as those for realisation 6, but was developed with a slightly lower, albeit sufficient resolution to prevent artificial fragmentation (Truelove criterion satisfied). A comparison of the plots on the two panels of Fig. 9 suggests little qualitative difference between the outcomes from the respective realisations. We must, however, point out that while the satisfaction of the resolution criterion defined by Eqn. (10) above is sufficient to ensure there is no artificial fragmentation, as is indeed seen in this work, it is unlikely to be sufficient to obtain convergence in calculations of energy and momentum of the collapsing gas (Federrath \emph{et al.} 2014). In fact, below we compare the sink-formation timescales and the magnitude of gas-velocity in the radial direction within the post-collapse filament.

Evidently, cores  in the post-collapse filament indeed, appear like beads on a string. However, we note that not all density perturbations had collapsed to form cores when calculations were terminated. A few more would be expected to form in this filament, but the extant purpose of demonstrating a Jeans-type fragmentation of this post-collapse filament is served with the set of cores that can be seen on either rendered image. Finally, shown in Figs. 11 and 12 are the accretion histories for cores that form in realisations 6 and 7, respectively. There is not much difference between the timescale of fragmentation, as reflected by the epoch at which the first core appears in either realisation. Although, the formation of cores after the first one, in realisation 7 is somewhat delayed and form over a timescale of ~10$^{5}$ yrs, as against those in case 6 which form relatively quickly after the first core appears. However, it is likely that a more significant difference between timescales would be visible for a higher harmonic. Also evident from the comparative plots shown in Fig. 11 is the absence of convergence in the timescale of sink-formation and the mass of sink-particles in the two realisations likely due to inadequate numerical resolution as was discussed in \S 3.3. Crucially, the number of sink particles remains the same irrespective of the choice of resolution which, it was argued previously is sufficient to prevent artificial fragmentation even for the realisation with the lowest realisation in this work. Also, the accretion histories for sink-particles in either case are qualitatively similar.

\begin{figure}
   \centering
   \includegraphics[angle=270,width=0.5\textwidth]{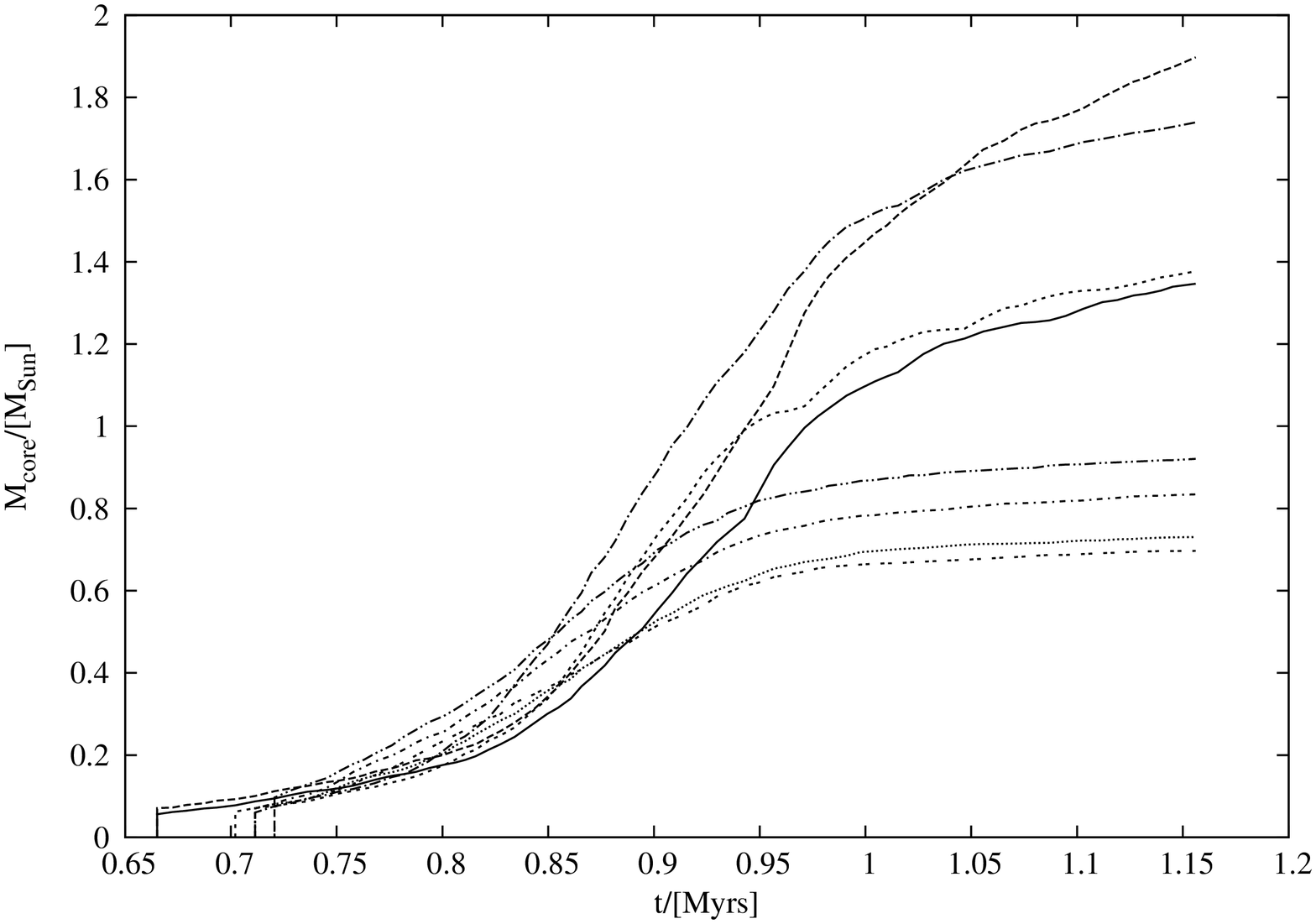}
    \includegraphics[angle=270,width=0.5\textwidth]{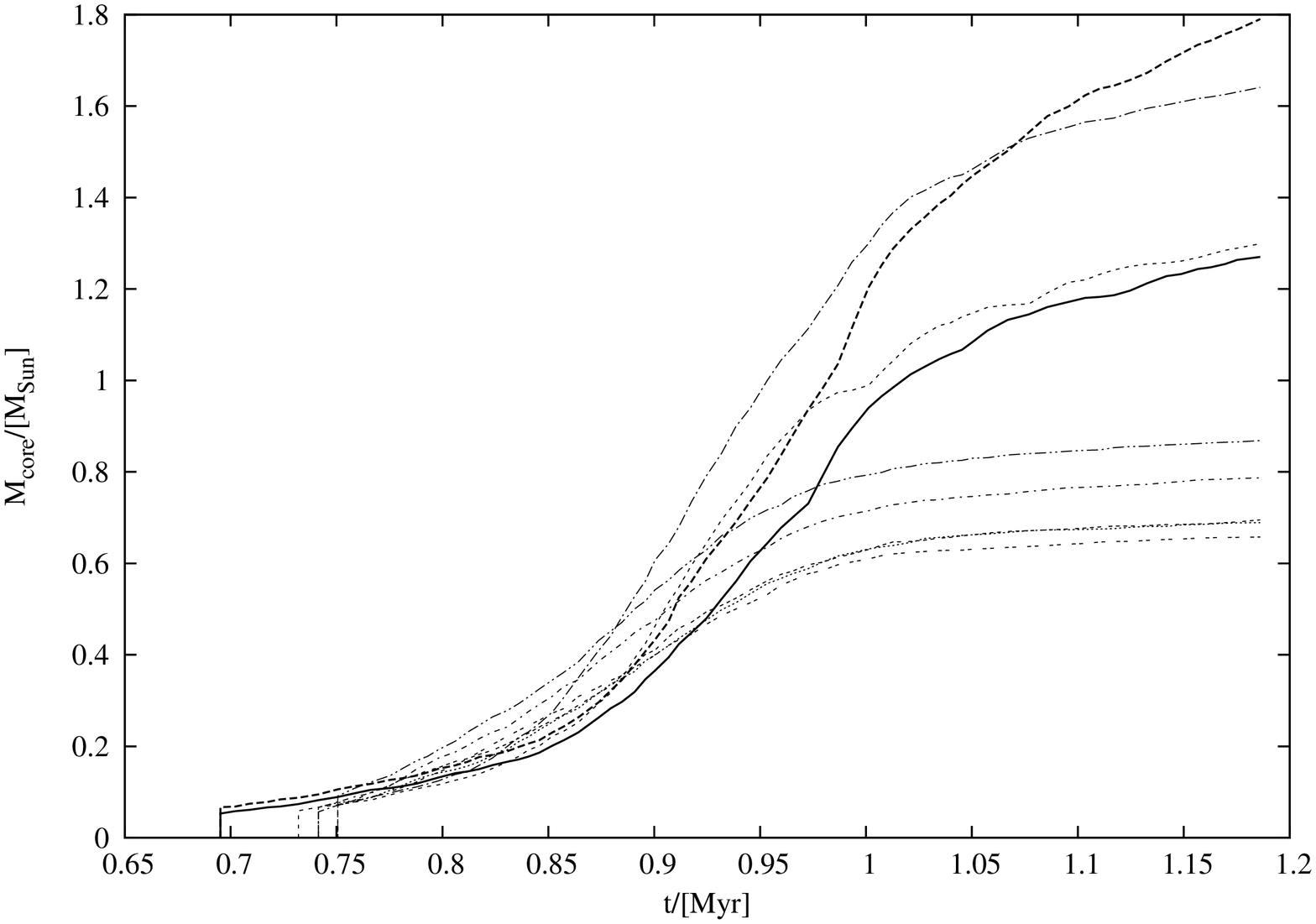}
      \caption{Same as the plot in Fig. 8, but now for the cores that form in the post-collapse filament in realisation 4 (\emph{upper-panel}) and 6(\emph{lower-panel}). In the former, the realisation with relatively lower-resolution, the sink-formation begins a little earlier and sinks are a little more massive than those in the latter realisation, one of those that had the  highest resolution in this work. }
         \label{FigVibStab}
   \end{figure}

\begin{figure}
   \centering
   \includegraphics[angle=270,width=0.5\textwidth]{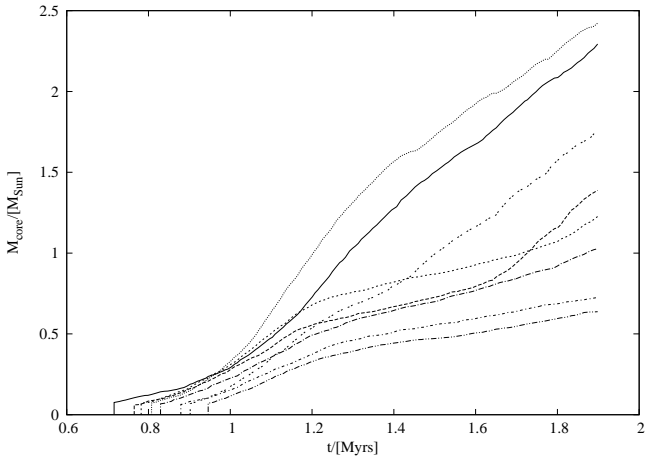}
      \caption{Same as the plot in Fig. 10, but now for the cores that form in the post-collapse filament in realisation 7.}
         \label{FigVibStab}
   \end{figure} 

\subsection{Case 2 (Serial nos. 3 and 8 in Table 1; $\frac{P_{ext}}{k_{B}}$ = 2.4$\times\ 10^{4}$ K cm$^{-3}$)}
In this second case we raised the gas temperature such that the initial cylindrical distribution of gas was rendered sub-critical. 
 We performed one realisation for this choice of $P_{icm}$ and repeated it with a higher resolution, numbered 8 in Table 1. Here we discuss this latter realisation. In contrast to the simulations discussed under case 1, the radial collapse of the cylinder could not be sustained as it became to be squashed by the confining pressure into a spheroidal globule that was in approximate equilibrium with the external medium. The other notable feature being, this spheroidal globule was assembled on a relatively longer timescale, $t\sim$ 2.5 Myrs; in comparison to that observed in the simulations tested under case 1.  As with the realisations discussed earlier under Case 1, shown in Fig.13 is the stability factor, $S(r)$, for this realisation. This is plot is similar to the ones shown earlier in Figs. 3 and 6 in the sense that gas within the cylindrical cloud always remains sub-critical as $S(r<R) < 1$. However, it differs from the former plots in one respect; the magnitude of $S(r)$, in the outer regions, away from the central axis of the globule, shows a significant fall relative to that for the initial configuration. This, as we will demonstrate later, is because gas in the outer regions is significantly warmer than that observed in realisations grouped under Case 1.

\begin{figure}
   \centering
   \includegraphics[angle=270,width=0.5\textwidth]{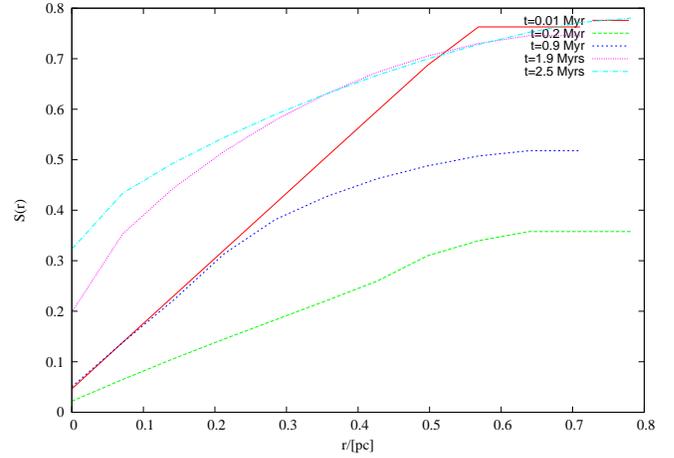}
      \caption{The radial variation of the stability factor, $S(r)$, calculated at different epochs for the cylindrical distribution of gas in realisation 8. As was seen in Figs. 2 and 5 for realisations grouped under Case 1, in this realisation as well the magnitude of $S(r)$ remains unchanged and gas always remains sub-critical even as the initial cylindrical distribution is squashed into a spheroidal globule.}
         \label{FigVibStab}
   \end{figure}

\begin{figure}
   \centering
   \includegraphics[angle=270,width=0.5\textwidth]{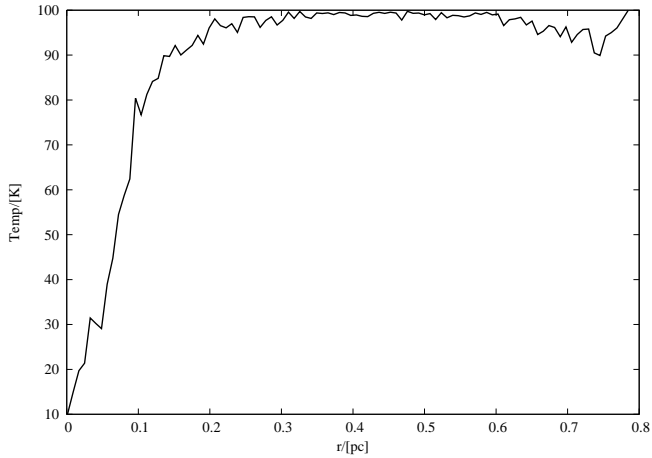}
      \caption{Radial distribution of the density averaged gas temperature within the post-collapse globule in realisation 8 at the time of termination of calculations ($t=2.5$ Myrs).}
         \label{FigVibStab}
   \end{figure}

\begin{figure}
   \centering
   \includegraphics[angle=270,width=0.5\textwidth]{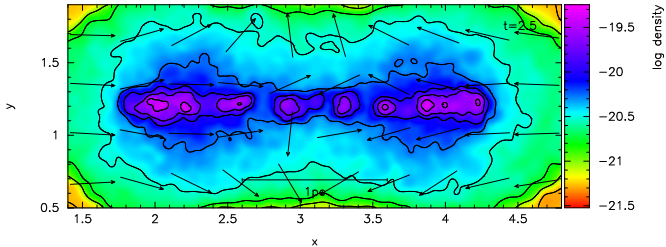}
      \caption{A rendered density image showing a projection of the mid-plane of the elongated globule in simulation 8. Time in the top left-hand corner of the image is marked in Myrs ($t$ = 2.5 Myrs ). }
         \label{FigVibStab}
   \end{figure}

\begin{figure}
   \centering
   \includegraphics[angle=270,width=0.5\textwidth]{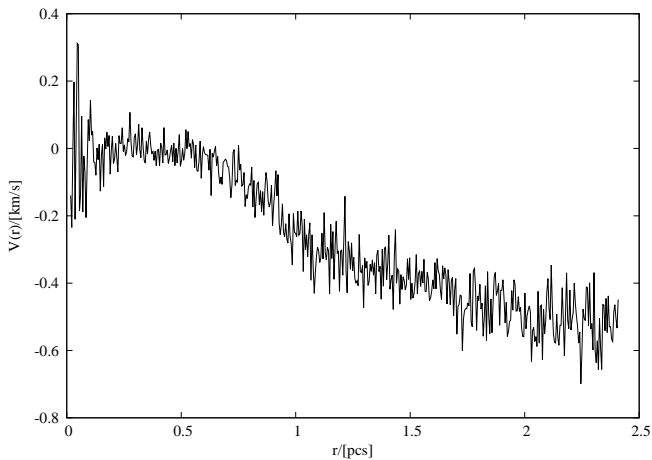}
      \caption{Gas within the post-collapse globule shown in Fig. 14 exhibits a weak velocity field. The inwardly directed gas in the outer regions of the globule as it is being squashed has negative velocity on this plot. }
         \label{FigVibStab}
   \end{figure}

 Shown in Fig. 14 is the density averaged gas temperature within this globule and was made at the time of terminating the calculations in this case ($t\sim$ 2.5 Myrs). From this plot it is clear that though the interiors of this globule are indeed cold, comparable to the central regions of the dense filament in case 1, gas away from the central axis is significantly warmer. The relatively large gas temperature renders it gravitationally sub-critical and so, it is unable to collapse radially to assemble a thin dense filament along the axis of the initial distribution. Instead, it begins to be squashed from either side and proceeds to form an elongated spheroidal gas-body as can be seen in the rendered density image shown in Fig. 15. Overlaid on this image are density contours that readily reveal evidence of sub-fragmentation to form smaller cores along the cold central region of the globule; velocity vectors on this image indicate the direction of local gas-flow within the elongated globule. 

The magnitude of this velocity field has been plotted in Fig. 15. Note that the magnitude of this velocity field is consistent with that derived for typical cores (e.g. Motte \emph{et al.} 1998; Jijina \emph{et al.} 1999). However, the local direction of the velocity field is unlikely to reveal much about the boundedness of a core which reinforces our suggestion made in an earlier work related to modelling starless cores. There it was shown, a core could remain starless despite exhibiting inwardly pointed velocity field which usually, is indicative of a gravitational collapse (Anathpindika \& di’Francesco 2013). Formation of sub-structure within this globule can be identified with the aid of density contours. This indicates the onset of sub-fragmentation. The fragments, at this epoch though, have not reached the density threshold for cores adopted in this work.

\section{Discussion}
Prestellar cores usually appear embedded within dense filamentary clouds. Understanding the morphology of these clouds is therefore crucial towards unravelling the details of the star-forming process. In the present work we explore the possibility of formation of prestellar cores along the length of a filamentary cloud. To this end we developed hydrodynamic simulations starting with a cylindrical distribution of gas and separately studied its evolution when it was gravitationally sub-critical and super-critical. We identify two possible modes of evolution : one, 
when the initial cylindrical distribution of gas is either marginally super-critical($S(R)\gtrsim$ 1), or even critically stable($S(R)\sim$ 1), a thin dense filament forms as a result of radial collapse of the initial distribution of gas. The post-collapse filament continues to accrete gas during in-fall and prestellar cores form along the length of this filament via a Jeans-like instability. In literature this mode has been identified as a tendency to form a spindle (e.g. Inutsuka \& Miyama 1998); and two, when the initial distribution of gas is gravitationally sub-critical($S(R)< $1), it tends to get squashed and forms a spheroidal globule that is in approximate pressure equilibrium with its confining medium. The relatively cooler regions of this globule show evidence for further fragmentation into smaller cores. Similar examples of fragmentation within larger cores to form smaller ones has been identified in for instance, the Serpens North (Duarte-Cabral et al. 2010), or within a number of infrared dark clouds (e.g. Wilcock et al. 2011, 2012).

Previous numerical work by a number of authors on the formation of filamentary clouds, and their stability against self-gravity, has culminated in at least three possible scenarios : (i) Filamentary clouds are often seen in magneto-/hydrodynamic simulations of turbulent gas. These clouds are believed to be generated due to interaction between turbulent gas, followed by enhancement of self-gravity once the gas becomes sufficiently dense (see for e.g. Klessen \& Burkert 2000; Klessen et. al. 2000; Federrath \& Klessen 2013). This model favours filament formation via interaction between turbulent flows so that there is no global collapse; dense filamentary clouds instead, appear locally (e.g. Federrath et al. 2010; Peters et al. 2012), (ii) Heitsch et al. (2010, 2013), have demonstrated that filamentary clouds could also form within a self-gravitating MC, and (iii) fragmentation of gas-sheets is also a possible mode of formation for elongated clouds. This has been demonstrated in both, the simple case of an isothermal sheet (e.g. Schmid-Burgk 1967; Myers 2009), as well as in the one confined by shocks (e.g. V{\' a}zquez-Semadeni et al. 2007; Anathpindika 2009; Heitsch 2010).

 Although there is little doubt about the fact that dense filaments are a result of fragmentation of larger clouds and that prestellar cores usually form within such filaments like beads strung on a wire, the process(es) leading to the formation of a thin filament are somewhat unclear and is still a matter of debate. Heitsch (2013), for instance has suggested that filaments, during their
formation, do exhibit density enhancement in their central regions, but this density enhancement is not associated with a corresponding reduction in the filament-width. Smith et al. (2014), partially agree with this conclusion but have suggested that dense filaments
are likely to have a static density-profile, one that does not vary with time. In other words filament-formation is probably a one step process and once created, dense filaments remain as they are. In simulations grouped under case 1 of this work we have demonstrated the formation of a thin dense filamentary cloud via radial collapse of the initial cylindrical distribution of molecular gas. Peters \emph{et al.}(2012) argue that the magnitude of the externally confining pressure probably determines if whether a dense filament could possibly form. These authors have suggested that a relatively small magnitude of the external pressure is likely to support the formation of a dense filament. On the contrary, a filamentary distribution of gas is more likely to end up as a spheroidal globule when the magnitude of confining pressure is relatively large. However, the possible cause(s) for such occurrence is(are)
not clear from their work. Conclusions drawn from the simulations presented in this work are broadly consistent with those of Peters \emph{et al.}(2012), but with an added qualification about the gravitational state of the distribution of gas that precedes a dense filament. This leads us to the next question, that about the propensity of a cylindrical cloud to collapse radially and assemble a dense filament. We observe that when the initial distribution of gas is gravitationally sub-critical, apart from a higher magnitude of confining pressure, the gas does not collapse in the radial direction, instead, it forms a spheroidal globule. We therefore believe, the gravitational state of the initial distribution of gas is likely to hold the key to its future evolution.

We argue, a radial  collapse leading to the formation of a dense filament is possible when the gas is at least critically stable($S(R)\sim$ 1), initially. Interestingly, the mass line-density for the post-collapse filament in this case exceeds its maximum value, $(M_{l})_{max}$, for stability under the assumption of a uniform temperature of ~10 K, as is indeed seen in the interior of the filament; though gas in the wings of the filament is warmer. Under this assumption of isothermality and uniform density the filament could be described as gravitationally \emph{super-critical} (Fischera \& Martin 2012). Although the two sets of simulations discussed in this work demonstrate that an initially sub-critical cylindrical volume of gas cannot be induced to collapse in the radial direction, plots shown in Figs. 2 and 5 suggest that gas in a radially collapsing cylindrical cloud does not become gravitationally super-critical ($S(r<R) <$ 1). In other words, the mass line-density, $M_{l}$, within the cylinder never exceeds its maximum value, $(M_{l})_{max}$, at any radius within the cylinder. Simulations grouped under this case(Case 1) also demonstrate that a filament need not be in radial free-fall in order to form prestellar cores along its length. Interestingly, evidence for global in-fall has recently been reported in the filamentary cloud DR21 (Schneider et al. 2010), while Palmeirim et al. (2013) and Andr´e et al. (2014) also report observations of some filamentary clouds that exhibit radial collapse. Inutsuka \& Miyama (1997), by assuming super-critical initial conditions and an isothermal gas had demonstrated such a collapse. However, this latter set of initial conditions would always be predisposed to radial collapse leading to the formation of a thin dense filament. Despite their propensity to collapse radially, globally super-critical filaments are unlikely to be in free-fall. The difficulty with the idea of a radially free-falling filament is that that filament would collapse rapidly to form a thin line and it would be a challenge to explain the formation of cores in it. It is therefore difficult to reconcile the suggestion of free-falling filaments. We will revisit this point in the context of observations of filaments in the following subsection. 

That filamentary clouds are unlikely to experience a free-fall collapse can simply be shown by deriving an expression for the radial component of gravitational acceleration, $g_{r}$, within a typical filamentary cloud. For the sake of argument, it would be safe to adopt a Plummer-density distribution for gas within a filamentary cloud so that the radial distribution of thermal pressure within this cloud may be written as, 
\begin{equation}
\frac{dP(r)}{dr}\ =\ a_{0}^{2}(r)\frac{d\rho(r)}{dr} + \rho(r)\frac{da_{0}^{2}(r)}{dr}.
\end{equation}
From the Plummer-density profile we have,
\begin{equation}
\frac{d\rho}{dr}\ =\ \frac{-2\rho_{c}}{r_{c}^{2}}\cdot\frac{r}{(1 + (\frac{r}{r_{c}})^{2})^{2}}.
\end{equation}
Combining Eqns. (12) and (13), we wind up with an expression for the gravitational acceleration as,
\begin{equation}
g_{r}(r)\ \equiv\ \frac{-1}{\rho(r)}\Big(\frac{dP(r)}{dr}\Big)\ =\ \frac{2\rho_{c}a_{0}^{2}(r)}{\rho(r)r_{c}^{2}}\cdot\frac{r}{(1 + (\frac{r}{r_{c}})^{2})^{2}} - \frac{da_{0}^{2}(r)}{dr}.
\end{equation}
Then in the limit of $r\rightarrow r_{flat}$, Eqn. (14) reduces to
\begin{displaymath}
g_{r} \sim \frac{a_{0}^{2}(r=r_{flat})}{4r_{flat}} > 0,
\end{displaymath}
which suggests that filamentary clouds are unlikely to have a true super-critical state, i.e., filaments are unlikely to be in free-fall even if they do collapse radially. We have shown that prestellar cores are likely to form via a Jeans-type fragmentation of the filamentary cloud and have demonstrated this by developing simulations with and without initial perturbations to the density field. This scenario is consistent with the argument presented by Freundlich \& Jog (2014). We also note that a Plummer-like profile ($p$ = 2), fits the radial density distribution of the post-collapse filament very well when gas temperature is calculated by accounting for gas-cooling. On the other hand, the steeper Ostriker-profile ($p$ = 4), fits the radial density profile for the isothermal filament.

Furthermore, the typical radius of the post-collapse filament in our simulations at the epoch when calculations were terminated, is on the order of $\sim$0.1 pc. Both these findings are consistent with corresponding values derived for filamentary clouds in the Gould-Belt (e.g. Arzoumanian \emph{et al.} 2011; Malinen \emph{et al.} 2012; Palmeirim \emph{et al.} 2013; Andr{\' e} \emph{et al.} 2014). Although a few handful number of exceptions have been reported with relatively steep density profiles in outer regions and have slopes in excess of 2, e.g. the B211/3($p$ = 2.27) filament in the Taurus {\small MC} that also has a relatively small radius of $\sim$0.04 pc (Malinen \emph{et al.} 2012; Palmeirim \emph{et al.} 2013). However, it remains to be seen if these post-collapse dense filaments would continue to self-gravitate and acquire even smaller radii; investigation into this question is best deferred for a future work. We also note that purely hydrodynamic simulations such as the ones discussed in this work succeed in reproducing the typical Plummer-like radial density profile for filamentary clouds (also see Smith \emph{et al.} 2014). This is contrary to some of the earlier suggestions about the propensity of formation of such filaments in magneto-hydrodynamic simulations (e.g. Fiege \& Pudritz 2000; Tilley \& Pudritz 2003; Hennebelle 2003). It appears that the slope of the radial profiles of filamentary clouds is unlikely to be influenced by the presence/absence of the magnetic field.

Also, unlike much of the earlier work we have also derived the radial distribution of gas density and temperature for the post-collapse filament and showed that is indeed consistent with that derived observationally for typical filamentary clouds observed in nearby star-forming regions. An interesting point to which attention must be drawn is about the temporal evolution of the density profile of filamentary clouds. Radial density profiles shown in Figs. 2 and 5 demonstrate that the collapse of the initial cylindrical distribution of gas is accompanied with a rising central density. This is consistent with the finding of Heitsch (2013). However, we also observe that the width of the filament shrinks during the process which is inconsistent with the other finding by the same author and Smith \emph{et al.} (2014). As can be seen from these plots, the problem could be circumvented by adopting a value of the $r_{flat}$, the point where the density-profile turns-over, smaller than where it is actually seen to lie. Recchi \emph{et al.} (2014) have argued that rotation tends to stabilise a filamentary cloud against self-gravity and leads to a density-profile shallower than the Ostriker-profile. On the contrary, we have shown that even a simple prescription of gas-cooling can indeed reproduce a Plummer-like density distribution for the post-collapse filamentary cloud.

On the other hand, upon raising the magnitude of external pressure, $P_{icm}/k_{B}$, such that the initial distribution of gas was rendered sub-critical (because the condition of pressure balance at the gas-{\small ICM} interface demands that gas temperature be raised), we observed, such a volume of gas was unable to sustain a collapse in the radial direction. Instead, after initially shrinking along the radial direction, gas was squashed laterally and therefore tended to rebound. Confined by external pressure, this gas then assembled a pressure-supported spheroidal globule on a time-scale larger than that for simulations discussed above. Interestingly, Arzoumanian et al. (2013) have recently identified filamentary clouds with a greater internal pressure in the IC5146, Aquila and the Polaris molecular clouds. These filaments have a relatively large velocity dispersion and therefore, are gravitationally unbound. Some of these in fact, exhibit signs of lateral expansion with little evidence for core-formation along the filament-axis as is usually envisaged in the \emph{beads-on-string} analogy. On the basis of results obtained from our simulations we suggest, factors controlling the efficiency with which molecular gas cools would determine the propensity to form dense filamentary clouds which is simply a restatement of the stability argument presented in \S 2. A larger thermal pressure could have an impact on the molecular chemistry of a cloud and therefore, on the efficiency with which it could possibly cool. However, in the absence of a complete treatment of the molecular chemistry we cannot address this issue in the present work.

\subsection{Possible implications for observational studies of filamentary clouds}
Irrespective of whether the initial cylindrical distribution of gas ends up as a dense filament, we observe that gas in the collapsing cylinder remains gravitationally sub-critical, characterised by the stability factor, $S(r< R) < 1$, at all epochs (see Figs. 3 and 6). This suggests, the gravitational state of the gas remains unaltered over the course of its evolution. Of particular interest is the implication of this finding for dense filaments. We have seen that the formation of a dense filament is the result of a radial contraction of the initial cylindrical distribution of gas. This contraction is associated with a rise in the central density along the axis of the cylinder whence a filamentary cloud is assembled. Although this is our observation in the simulations discussed here, we note that the radially collapsing gas is not in a free-fall and is therefore characterised by $S(r<R)<1$, at all radii within the volume of gas. This sequence supports the hypothesis that filamentary clouds are probably assembled via gravitational contraction during which mass is steadily accreted by the centrally located filamentary cloud. This scenario is further reinforced by observational findings of accreting filaments found in Taurus and Cygnus X (Goldsmith \emph{et al.} 2008; Nakamura \& Li 2008; Schneider \emph{et al.} 2010).

\begin{figure}
   \centering
   \includegraphics[angle=270,width=0.5\textwidth]{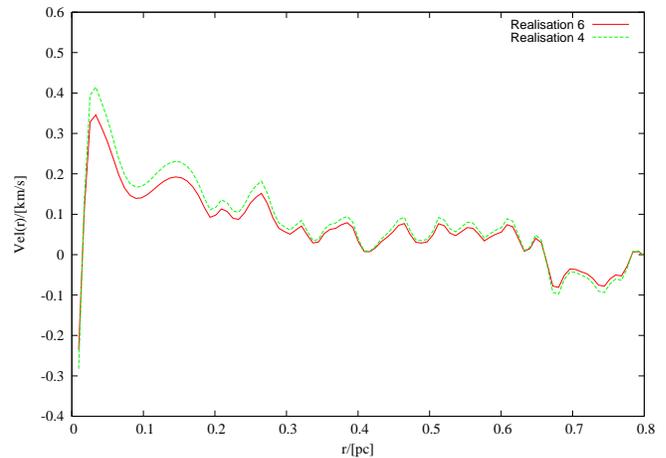}
      \caption{Same as the plot shown in Fig. 16 but now for simulations 4 and 6 grouped under Case 1. As usual, a negative velocity indicates inwardly moving gas. Although the plots are qualitatively similar, the resolution of the simulations developed here is insufficient to achieve convergence in the magnitude of velocity of the collapsing gas. (See \S 3.3 for a discussion about the resolution).} 
         \label{FigVibStab}
   \end{figure}

A comparison of the radial velocity field in the two cases discussed in this work would be instructive. For Case 1 we take the sixth realisation, one of the two in this set with the highest resolution, as a representative calculation. The radial velocity field in the post-collapse filamentary cloud has been shown in Fig. 17.  We note that the magnitude of velocity is relatively large in the central region and peters-off in the wings of the filament. In fact, the radial distribution of gas velocity can be approximated by a power-law of the type, $V(r)\propto r^{-0.4}$, which is very close to that for a bound object ($V(r)\propto r^{-0.5}$). A similar plot for simulation 8 under Case 2 was shown previously in Fig. 16. The difference between these plots is evident; in this latter case, the gas close to the centre is moving radially outward where as that in the outer regions is transonic and moving inwards with approximately constant velocity (sound speed $\sim$0.55 km/s in this region), as a consequence of being squashed. The findings reported here from simulations in Case 1 are consistent with those for  typical filamentary clouds (see for e.g. Arzoumanian \emph{et al.} 2013).

Filaments of the type seen in Case 1 are conventionally described as gravitationally bound. This brings us to the next question - as to whether gravitationally bound filaments are also likely to experience  a free-fall collapse. For the purpose of illustration we consider the examples of four filamentary clouds. (i) IC5146 in the Cygnus region. This filament with a mass line-density, $M_{l}\sim$ 152 M$_{\odot}$ pc$^{-1}$, which is an order of magnitude in excess of the maximum mass line-density, $(M_{l})_{max}$, evaluated at 10 K, and therefore should be expected to be in a free-fall collapse in the radial direction. However, it seems, this is not really the case as the filament does not appear to be in radial free-fall (Arzoumanian \emph{et al.} 2011; Fischera \& Martin 2012). (ii) The filamentary cloud {\small DR}21 is the next example. This is a massive filament that has two sub-filaments that are gravitationally super-critical and show signs of in-fall, but no apparent signs of radial free-fall (Schneider \emph{et al.} 2010). This filament appears more akin with the outcome of simulations in Case 2 of this work. (iii) Next, we consider the example of the Perseus {\small MC}. It has a mass line-density, $M_{l}\sim$ 50 M$_{\odot}$ pc$^{-1}$ to 100  M$_{\odot}$ pc$^{-1}$, a good factor of 3-5 larger than the maximum mass line-density, $(M_{l})_{max}$,  for the deduced isothermal temperature of $\sim$12 K (Hatchell \emph{et al.} 2005). Again, a direct comparison of $M_{l}$ against $(M_{l})_{max}$ for the Perseus {\small MC} would suggest that the filament is gravitationally bound and must collapse radially. However, there is no corroborative evidence to suggest that the filament is in free-fall though several cores have been detected along its length. In fact, estimates of line-width of the optically thick C$^{18}$O emission suggests that that filament is probably thermally supported. (iv) Finally, we consider the example of the Serpens {\small MC}, recent observations of which, have revealed evidence of radially infalling gas onto the dense filaments in this cloud. Yet, there is no evidence to suggest that these filaments are in free-fall, however, prestellar cores have been detected along their length and appear to have formed on the scale of the local Jeans length (Friesen \emph{et al.} 2014). Results from our simulations grouped under Case 1 corroborate these observations.

It therefore appears to us that the condition, $S(r=R) \ge 1$, i.e., $\frac{M_{l}(r=R)}{(M_{l}(r=R))_{max}}\ge1$, is probably necessary to induce radial collapse in a cylindrical volume of gas.
Such a volume of gas could become super-critical by accreting mass from its surroundings (Heitsch \& Hartmann 2014). However, this aspect of the problem is beyond the scope of present investigation and therefore best left for a future work. Significantly though, this collapsing gas is unlikely to be in free-fall and $S(r < R)< 1$ which is consistent with the suggestion of McKrea(1957); see also Toci \& Galli (2014). In fact, if indeed this gas is in radial free-fall, filamentary clouds would rapidly end up in a thin line and formation of cores in such a collapsing cloud would be difficult to reconcile. On the contrary, a steady in-fall, as gas within the collapsing filament is allowed to cool appears to be a more promising mode of evolution of filamentary clouds where gas is initially at least critically stable. As we have seen in the simulations grouped under Case 1, such filaments can indeed form cores along their length via a Jeans-like fragmentation. This evolutionary scenario also circumvents the uneasy question as to why filaments with extremely high column densities, on the order of a few times 10$^{22}$ cm$^{-2}$, are not found. The likely solution perhaps is that filaments are not in radial free-fall. Those that are gravitationally bound, shrink in the radial direction and eventually acquire a thermally supported configuration.

If indeed filaments were to become pressure-supported during their evolutionary sequence, it would also explain why filaments across star-forming regions tend to have comparable widths, on the order of $\sim$0.1 pc (e.g. Arzoumanian \emph{et al.} 2011). At 10 K, the temperature typically found in filamentary clouds and an average density $\sim$10$^{-19}$ g cm$^{-3}$, a representative density for prestellar cores, the thermal Jeans length, $L_{Jeans}\sim$ 0.075 pc, which is comparable to the typical radius of filamentary clouds. We make a similar observation in simulations grouped under Case 1. For instance, Figs. 2 and 5, where the radial density profile for the post-collapse filament in simulations 1 and  2 have been plotted, demonstrate that the filament radius, $r_{flat}\sim\ L_{Jeans}$, at the epoch when calculations were terminated in respective simulations. Although, we note that magnetic field could also have a key role to play in the evolution of filamentary clouds. The point though, is beyond the scope of this article.

\section{Conclusions}
In this work we have demonstrated that an initial cylindrical distribution of molecular gas can indeed collapse radially to assemble a thin dense filament along its axis when the magnitude of confining pressure is relatively small such that gas is initially at least critically stable. In this case a radial collapse ensues and the gas is able to cool on a relatively short timescale. Equivalently, it may also be argued that a collapse of this kind is likely when the line mass exceeds its critical value required to maintain stability. However, there is a caveat to this argument. This argument is valid only if gas is assumed to be isothermal and has uniform density as was the case with our initial cylindrical distribution of gas. For a typical dense filament this argument cannot be applied since its density and temperature distribution is far from uniform. We argue, a filamentary cloud even if gravitationally bound, is unlikely to be in radial free-fall,  instead it is likely to contract in the radial direction and consequently, the density along its central axis steadily rises as gas is accreted on to it. We argue that this evolutionary sequence of a typical filamentary cloud could possibly explain - (i) the dearth of filamentary clouds that have extremely high extinction, and (ii) why the width of filamentary clouds shows little variation across star-forming regions. We suggest that filamentary clouds are likely to be thermally-supported and their radii(width) are likely to be on the order of the local thermal Jeans length (twice the local thermal Jeans length). A Plummer-like density profile appears to fit very well the radial distribution of gas density in the post-collapse filament so that magnetic field may not be necessary to generate filaments with a relatively shallow slope, as has been suggested in the past (e.g. Fiege \& Pudritz 2000). We also demonstrate that prestellar cores are likely to form via a Jeans-like instability  while it is accreting gas. It also appears, cores are likely to form along the length of a dense filament if it is at least (gravitationally)critically stable. Processes via which filaments acquire mass from their parent cloud are therefore likely to hold the key to determining the efficiency of star-formation (see also Tafalla \& Hacar 2014)
 
On the other hand,  when the magnitude of the confining pressure was increased such that the initial distribution of gas was rendered gravitationally sub-critical, we observed that a radial collapse was unsustainable. Instead, the gas distribution was squashed from either side and demonstrated a propensity to expand laterally. The result was an elongated spheroidal globule which then showed signs of sub-fragmentation. Such objects are more likely to form out of dynamic interactions between gas flows within molecular clouds. The upshot therefore is that potential star-forming dense filaments are likely to be found in molecular clouds where ambient conditions contrive to render the volume of gas immediately preceding the dense filament critically stable, at least (i.e. mass line-density approximately equal to its maximum stable value). The possible effect of the ambient conditions on the chemistry within a molecular cloud must directly impact the gas thermodynamics and therefore, the dynamical stability of that cloud. We will examine this issue in a future contribution. 

\begin{acknowledgements}
 The authors thank an anonymous referee for a helpful and prompt report. J.F. acknowledges support from the Indo-French Centre for the Promotion of Advanced Research (IFCPAR/CEFIPRA) through a Raman-Charpak fellowship over the course of which this project was conceived.
\end{acknowledgements}




\end{document}